%% file: iclr2019_conference.tex
\documentclass{article} 
\usepackage{iclr2019_conference,times}

\input{math_commands.tex}

\usepackage{hyperref}
\usepackage{url}
\usepackage{amsthm}
\usepackage{amsmath}
\usepackage{amssymb}
\usepackage{booktabs}
\usepackage{textcomp}
\usepackage{dsfont} 

\usepackage{array,multirow,graphicx}

\theoremstyle{definition}

\title{\center{Unsupervised Community Detection with Modularity-Based Attention Model}}

\author{Ivan Lobov, Sergey Ivanov \\
Criteo\\
Paris, France \\
\texttt{\{i.lobov,s.ivanov\}@criteo.com} \\
}

\iclrfinalcopy 
\begin{document}

\maketitle

\begin{abstract}
In this paper we take a problem of unsupervised nodes clustering on graphs and show how recent advances in attention models can be applied successfully in a "hard" regime of the problem. We propose an unsupervised algorithm that encodes Bethe Hessian embeddings by optimizing soft modularity loss and argue that our model is competitive to both classical and Graph Neural Network (GNN) models while it can be trained on a single graph.
\end{abstract}

\section{Introduction}
Community detection on graphs is a task of learning similar classes of vertices from the network's topology. It is one of the central problems in data mining and has found numerous applications in sociological studies (\cite{sociological}), DNA 3D folding (\cite{dna}), product recommendations (\cite{recommendation}), natural language processing (\cite{nlp}) and more. It is not surprising that many fundamental results have been obtained in recent years that shed the light on our understanding of the solvability of this problem in general. Specifically, precise phase transition and fundamental limits exist for one of the central graph generation models to test community detection algorithms, the so-called Stochastic Block Model  SBM($n$, $p$, $W$) and its symmetric variant \textit{Symmetric Stochastic Block Model}, SSBM($n$, $k$, $A$, $B$) (\cite{sbmrecent}). In this paper, we'll be dealing with a simpler case, SSBM.

The optimization of log-likelihood and modularity are equivalent in SSBM model \citep{newman2013spectral}. Hence, we propose to use a generalization of modularity that outputs the probability of each cluster for a node and is therefore differentiable for a neural network model. Our model is based on the recent advances in NLP where Transformer model (\cite{attentionneed}) with attention mechanism shows superior results. We adopt the encoder part of Transformer to transform initially obtained Bethe Hessian embeddings and then produce the probability of each cluster for each node to optimize with our loss function. 

Our contributions include:
\begin{itemize}
\item A new model with attention mechanism that optimizes a soft modularity loss function (and release the code to reproduce our results\footnote{https://github.com/Ivanopolo/modnet}). 
\item A comparative study of classical algorithms and recent supervised graph neural network and outline several advantages of the unsupervised model. 
\end{itemize}

\section{Related Work}

SBM models have seen a resurged interest recently due to the conjecture of (\cite{overlap}) that hypotheses phase transition for weak recovery case (the one we are interested in this paper) and the existence of the information-computation gap for 4 communities in the symmetric case. It was proved in (\cite{k21, k22, k23}) that efficient algorithms exists when there are 2 communities and signal-to-noise ratio is greater than one (KS threshold), while (\cite{k24}) shows that it is impossible to detect communities below this threshold for 2-community case. For more than 3 communities it was also proved that there are non-efficient algorithms that can weakly recover communities strictly below KS threshold (\cite{k41}). An extensive overview of the area can be found in \cite{sbmrecent}.

Graph neural networks have been recently applied to solve community detection problem with the current state-of-the-art LGNN model(\cite{lgnn}) designed to extract high-order node interactions via Non-backtracking operator. In parallel line of work, there are graph neural models based on attention mechanisms (\cite{gat, roting}) that have showed prominent results in other domains such as speech recognition and NLP. 

\section{SSBM Regimes}

Signal-to-Noise Ratio (SNR) for SSBM($n$, $k$, $a/n$, $b/n$) model is defined as:

\begin{equation}
    \text{SNR} = \frac{(a-b)^2}{k(a + (k-1)b)}
\end{equation}

It is known that for $k \ge 2$ and $\text{SNR} > 1$ it is possible to detect communities in polynomial-time (\cite{crossing}). Moreover, when $\text{SNR} \le 1$ and $k=2$ the weak recovery of SSBM model is not possible theoretically. However, it is also known that for $k \ge 4$ it is possible to weakly recover in the setup of $\text{SNR} > \alpha$, where $\alpha < 1$. This means that there exists a \textit{computational threshold} after which only information-theoretical algorithms can recover communities. This gap between what is possible to compute efficiently and what can be computed in general is known as \textit{information-computation gap}. In this work we compare two different regimes: one regime that lies within a computation threshold and is achievable by classical algorithms (associative case) and one regime for the information-theoretical gap (disassociative case).

\section{Approach}
\subsection{Loss}
If we know the generative model from which a graph is produced, it is possible to make statistical inference and evaluate the fitness of the estimated parameters using model's log likelihood. This approach assumes a hard labelling of the nodes and therefore requires some form of relaxation to obtain a differentiable loss function. It has been shown in \citep{newman2013spectral} that optimizing log-likelihood for the SSBM model is equivalent to the modularity optimization, a popular heuristic approach to graph clustering. And there exists \citep{havens2013soft} a generalization of modularity to soft labelling which is well differentiable and can be directly used for learning. \\
We take a soft modularity $M$ as our loss function:
\[
M=-tr(UBU^T)/||A||_{1}
\]
Where $U \in \mathbb{R}^{N \times C}$ is a matrix of probabilities of nodes attribution to clusters (the number of which is a model hyperparameter), $B=A-dd^T / ||A||_{1}$, $d$ is a vector of node degrees, $A$ is the adjacency matrix and $||A||_{1}=\sum{|A_{ij}|}$. \\
As the modularity might have multiple local optima, we found that it is beneficial if we can include additional information about the graph as a prior to steer the model in the right direction. We use a regularizer defined by:
\[
R=\left(\sum_{i}^{C} \left(\sum_{j}^{N} U_{ij}-\frac{1}{C} \right)^2 \right)
\]
Where $C$ is the number of communities in the graph. The regularizer forces the model to find communities of similar sizes as is expected in SSBM. \\

Our final loss is: 
\begin{equation}
\label{loss}
    \begin{cases}
        L_a=M+\lambda R, & \text{for associative communities} \\
        L_d=-M+\lambda R, & \text{for disassociative communities.}
    \end{cases}
\end{equation} 
We use negation of modularity loss $M$ as modularity is negative for disassociative case. The benefits of using this loss function are the following: we do not require explicit labels for learning, the loss is invariant under labels permutation and it is a consistent loss under SSBM with soft community partitioning.

\subsection{Model}
Our network takes as input initialized node embeddings and consists of three main blocks: projection into higher dimensions (done via a fully connected network with skip connections \citep{skip-connections} and batch norm \citep{batch-norm}), graph attention block and output prediction layer with a fully connected layer and a softmax.
\paragraph{Node Embeddings} We initialize node embeddings with eigenvectors recovered as part of the Bathe Hessian decomposition, a baseline algorithm described in Section \ref{baselines}.
\paragraph{Graph Attention} At every layer the attention is done over all neighbors of the node. The Encoder module is described in \citep{attention} which is a form of multi-headed attention on neighbors of a node. It is particularly well-suited for our initialization of the node embeddings; the key-value dot-products give the model access to the community structure estimated by the Bethe Hessian eigenvectors.
\subsection{Training}
Our model is an unsupervised model in a sense that it does not require any labels from the original graph. We train the model separately on each graph repeatedly trying to improve the soft modularity loss. Since the method does not require a larger training set, it significantly reduces the training time. \\
For all the experiments reported, we trained a model with 2 layers, 3 heads of attention, the size of all hidden layers equal to 48 and regularization parameter $\lambda=0.5$.

\section{Experiments}

\subsection{Datasets}
We generate two datasets SBM($n$, $k$, $a/n$, $b/n$) that correspond to associative and disassociative cases. For the disassociative dataset, we set $a=0$ and $b=18$, which means that there are no links between vertices of the same cluster. For the associative dataset, we set $a=21$ and $b=2$. All datasets have $n=400$ vertices and $k=5$ communities. The parameters for the disassociative case correspond to the information-theoretical gap, while for associative case they correspond to the regime when belief propagation can weakly recover the clustering efficiently. We generate 6000 graphs for training dataset and 1000 for validation. 

\subsection{Evaluation Metrics}
\textbf{Overlap} measures the intersection of the original assignment $y$ and the predicted assignment $\hat{y}$ across all possible permutations from the permutation group $S$ (\cite{overlap}):

\begin{equation*}
O(y, \hat{y}) = \max\limits_{\pi \in S_{\hat{y}}} \frac{(1/n)\sum\limits_u \delta_{y(u), \pi(\hat{y}(u)}) - 1/c}{1 - 1/c}
\end{equation*}

Where $n$ is the number of nodes in a graph, $c$ is the number of communities and $\delta_{u, v}$ equals one when $u = v$ and zero, otherwise. 

\textbf{Mutual information} is a similarity measure between two labels assignments $y$ and $\hat{y}$:
\begin{equation}
\label{MI}
I(y, \hat{y}) = \sum\limits_{i=1}^n\sum\limits_{j=1}^n \frac{y(i) \cap \hat{y}(j)}{n} \log n\frac{y(i) \cap \hat{y}(j)}{\vert y(i)\vert \vert \hat{y}(j)\vert}
\end{equation}
We use a normalized version (Normalized Mutual Information, hence NMI) of (\ref{MI}) by the arithmetic mean of the entropy for each assignment. \\
\textbf{Modularity} is a measure of fitness of our inference assuming the SSBM generative model (it differs by a constant factor from the log likelihood). It is positive for associative communities and negative for disassociative ones.

\subsection{Baseline Methods} \label{baselines}

We compare our model with a state-of-the-art GNN model, LGNN, and several classical baselines. \\
\textbf{Bethe Hessian}, introduced in \cite{bethe}, is a spectral approximation of Belief Propagation (BP), a classical algorithm for community detection, by employing the Bethe Hessian operator:
\begin{equation*}
    H = (r^2 - 1)\mathds{1} - rA + D,
\end{equation*}
where $r$ is the largest eigenvalue of non-backtracking operator $B$, $A$ is the adjacency matrix and $D$ is a diagonal matrix of degrees. We find the $k$ smallest algebraically eigenvectors of $H$ and find clusters using k-means algorithm. We found that its results are more stable and often better in practice than of BP.

\textbf{Louvain Modularity} \citep{louvain} is a popular baseline that greedily updates the clusters if the modularity is improved. Since it does not control the number of clusters and we often ended up with more clusters that in the original graph, we did not compute the overlap metric for it (which requires the same number of communities). We do not report results for the disassociative case as the algorithm cannot find communities in this setting by design, it can only merge nodes that are neighbors.

\textbf{LGNN} \cite{lgnn} is a supervised community recovery algorithm based on GNNs and that can learn operators of inter-graph node and edge connectivity similar to laplacian for nodes or non-backtracking operator for edges. 

\paragraph{GNN} \cite{kipf} In addition to the attention model, we also experiment with GNN architecture that produces network embeddings that are trained using our loss equation (\ref{loss}).

For our GNN and Attention models, we also have a choice of initialization of embeddings. We add results for Random and Bethe-Hessian (BH) initialization. 

\subsection{Empirical Results}

Results are presented in Tables \ref{table1} and \ref{table2}. Note that in disassociative case, the smaller value for modularity, the better; for all other metrics, the higher, the better. True labels denote the optimal values for the datasets. Among our 4 proposed algorithms (at the bottom), Attention model with Bethe-Hessian initialization is superior, showing that both the structure of the model and the initialization of embeddings improve the quality on all metrics. Overall, we can see that our modularity-based approach achieves competitive results for the associative and disassociative dataset even compared with supervised LGNN. Our model Attention-BH also achieves the top performance on modularity metric among all algorithms as soft modularity is explicitly present in the loss function. We can see that the loss function that we propose allows us to learn as good or even better fit for the generative model (SSBM), which shows that it can be efficiently used to find communities in a fully unsupervised way, learning only on the current graph.

\begin{table}[!htb]
    \begin{minipage}{.5\linewidth}
    
      \caption{Associative Dataset.}
      \label{table1}
      \vskip 0.15in
      \centering
        \begin{tabular}{| l | c | c | c  |}
              \hline
              Algorithm & Modularity & Overlap & NMI \\ \hline \hline
            True Labels         & 0.52 & 1.0& 1.0 \\ \hline \hline
            LGNN (supervised)             & 0.50 & 0.80& 0.67\\ \hline \hline
            Bethe Hessian         & \textbf{0.52}& \textbf{0.85}& \textbf{0.69} \\ \hline
            Louvain Modularity             & 0.48& N/A& 0.48\\ \hline \hline
            GNN-Random              & 0.47& 0.53& 0.48\\ \hline
            GNN-BH              & 0.49& 0.66& 0.61\\ \hline
            Attention-Random              & 0.42& 0.36& 0.27\\ \hline
            Attention-BH              & \textbf{0.51}& \textbf{0.78}& \textbf{0.67}\\ \hline
              \hline
        \end{tabular}
    \end{minipage}%
    \begin{minipage}{.5\linewidth}
      \centering
        \caption{Disassociative Dataset.}
        \label{table2}
        \vskip 0.15in
        \begin{tabular}{| c | c | c | c  ||}
              \hline
              Modularity & Overlap & NMI \\ \hline \hline
            -0.20& 1.0& 1.0  \\ \hline \hline
            -0.16 & 0.24 & 0.13\\ \hline \hline
             \textbf{-0.15}& \textbf{0.21}& \textbf{0.11} \\ \hline
            N/A& N/A& N/A\\ \hline \hline
            0 & 0.02 & 0.01\\ \hline
            -0.16 & 0.2 & 0.1\\ \hline
            -0.17 & 0.12 & 0.04\\ \hline
             \textbf{-0.18}& \textbf{0.22}& \textbf{0.11}\\ \hline
              \hline
        \end{tabular}
    \end{minipage}
\end{table}

\section{Conclusion}
In this work we propose a novel approach on how to recover communities in unsupervised way and conduct experiments comparing to state-of-the-art supervised neural models and unsupervised classical algorithms. Supervised models achieves the state-of-the-art results but with the price of having more parameters and longer training time. We believe it is interesting to combine LGNN architecture with an unsupervised modularity loss as potential future work.

\bibliography{iclr2019_conference}
\bibliographystyle{iclr2019_conference}

\end{document}

%% file: math_commands.tex
\usepackage{amsmath,amsfonts,bm}

\def\eqref#1{equation~\ref{#1}}

\def\1{\bm{1}}

\DeclareMathAlphabet{\mathsfit}{\encodingdefault}{\sfdefault}{m}{sl}
\SetMathAlphabet{\mathsfit}{bold}{\encodingdefault}{\sfdefault}{bx}{n}

